\documentclass[twocolumn,showpacs,preprintnumbers,amsmath,amssymb]{revtex4}
\usepackage{tabularx,graphicx}
\usepackage{color}
\usepackage{hyperref}
\hypersetup{
    colorlinks=true,
    linkcolor=blue,
    filecolor=magenta,      
    urlcolor=blue,
}
\begin{document}
\newcommand{\beq}{\begin{equation}}
\newcommand{\eeq}{\end{equation}}
\newcommand{\beqn}{\begin{eqnarray}}
\newcommand{\eeqn}{\end{eqnarray}}
\newcommand{\bmath}{\begin{subequations}}
\newcommand{\emath}{\end{subequations}}
\newcommand{\rb}{\bar{r}}
\newcommand{\bk}{\bold{k}}
\newcommand{\bkp}{\bold{k'}}
\newcommand{\bq}{\bold{q}}
\newcommand{\bkb}{\bold{\bar{k}}}
\newcommand{\br}{\bold{r}}
\newcommand{\brp}{\bold{r'}}
\newcommand{\vp}{\varphi}
\newcommand{\vs}{\vskip0.15in}
\newcommand{\red}{\textcolor{red}}

\title{Superconducting materials classes: introduction and overview}
\author{J. E. Hirsch$^{a}$, M.B. Maple$^{a}$, F. Marsiglio$^{b}$ }
\address{$^{a}$Department of Physics, University of California, San Diego, 
La Jolla, CA 92093-0319 \\$^{b}$Department of Physics, University of Alberta, Edmonton,
Alberta, Canada T6G 2E1
}

\begin{abstract} 
An introduction to and overview of  the contents of this Special Issue are given. 
32 classes of superconducting materials are discussed, grouped under the three categories ``conventional'', ``possibly   unconventional''
and ``unconventional'' according to the mechanism believed to give rise to superconductivity.
\end{abstract}
\maketitle 

\section{introduction}
 In this Special Issue we aim to   give a comprehensive overview of the superconducting materials  known to date. 
Superconducting materials were grouped into 32 different classes, and we invited recognized experimental leaders in each class, including
in many cases individuals who discovered a new class of superconductors, to contribute an article giving an overview of the properties of that class. We were fortunate to  get an excellent response.

By a ``class'' we mean a set of materials with common or closely related crystal structure, composition and physical properties, and hence presumably the members of
a class exhibit superconductivity driven by the same physical mechanism. There is at present no general understanding of the mechanism(s) giving rise to superconductivity in many of these classes.  The purpose of this Special Issue   is to put together in one place essential information on the multiple different classes of superconductors,  to facilitate comparison of the commonalities and differences in the physical properties of the different classes that may be playing a role in the superconductivity.

It is not uncommon that superconductivity researchers concentrate their efforts on  one or at most a small number
of classes of superconductors, and are unaware of relationships that may exist with other classes. 
We hope that  this compendium will facilitate making connections between the different classes of superconductors, thus helping researchers  identify which properties are linked to
 superconductivity of a given class and which are
not. We hope that this will contribute to the  ultimate goal of understanding what are the mechanisms of superconductivity that explain all the types of superconductors found in nature, and as a consequence aid in the search for new superconducting materials with desirable properties, particularly higher $T_c$'s.

The classes have been grouped into three categories: ``conventional superconductors'', ``possibly unconventional superconductors'' and ``unconventional superconductors''. For materials
in the first category, there is broad consensus   that they are described by conventional BCS-Eliashberg-Migdal theory of superconductivity, driven
by the electron-phonon interaction. For materials in the third category, there is broad consensus  that they are not described by the conventional theory, either because $T_c$ is too high or because some physical properties point to a different mechanism. However, there is no consensus on which new mechanism(s) explain the various different classes of unconventional superconductors. 
For materials in the second category, the evidence in favor of the conventional mechanism is mixed.  We asked the authors themselves to tell us in which of these three categories they felt that
their paper should be included and in most cases followed their 
recommendation. In cases where we did not receive input from the authors on this question we followed our best judgement.

We asked the authors  to  address for their class various normal and superconducting state properties, and in particular signatures of conventional or unconventional superconductivity.  The emphasis was to be on experimentally known properties, but since many of these measurements are motivated by an underlying theoretical framework, authors were also asked to   summarize theoretical ideas, such as band structure calculations, and proposed or currently accepted theoretical explanations, etc. We also asked the authors to address  commonalities and differences of their class with other classes that they felt may   be related.
Finally, we asked  authors to provide key references for each class including earlier reviews as sources of additional information for
the reader.

From the time superconductivity was discovered in 1911 until BCS theory was developed in 1957, many attempts at theories of superconductivity
were made \cite{theories}. The search during that period was for ``the'' theory of superconductivity, since it was   believed that 
a single theory would explain all the many superconducting elements and compounds already known at that time. After BCS theory was proposed, 
there were initially some suggestions that the BCS electron-phonon induced pairing mechanism may apply to 
most but not all superconductors \cite{rmp63,rmp64}, as reviewed in the contribution by Geballe
et al. \cite{geballe} in this volume. However, theoretical explanations within BCS theory were found for the anomalies that prompted these suggestions,
and by the time Parks' \cite{parks} influential treatise on superconductivity was published in 1969 and for about $10$ years thereafter, it was generally believed
that BCS-Eliashberg-electron-phonon  theory described all superconducting materials.

Nonetheless, some theoretical suggestions were made during the 60's and early 70's \cite{little,allender,ginsburg} 
that in some specially designed materials  
non-electron-phonon pairing mechanisms (``excitonic'') 
could give rise to superconductivity, potentially at higher temperatures, but no clear experimental evidence
for such materials was found. 

The situation began to change in the mid-1970's. The first material with a strong claim to be an ``unconventional superconductor''
was discovered by Sleight and coworkers  in 1975 \cite{sleight},  $BaPb_{1-x}Bi_xO_3$, with a surprisingly high $T_c$
($13K$) given its low density of states. A few years later, ``heavy fermion'' superconductors were discovered in 
1979 and organic charge transfer salts in 1980, both showing strong evidence for a non-s-wave order parameter. These were followed by the high $T_c$ cuprates in 1986, and many other classes of unconventional or possibly unconventional 
superconductors in the ensuing years. There is at this time no doubt in anybody's mind that  the conventional 
BCS-Eliashberg-electron-phonon theory of superconductivity is $not$
applicable to all superconductors.

It should not go without mention that Bernd Matthias, a prodigious researcher who discovered many new
 superconducting materials in the period 1950-1980 guided by empirical rules that he devised, had been vehemently
advocating the possibility of mechanisms other that the electron-phonon interaction to explain the superconductivity
of various materials during the 60's and 70's \cite{matthias_anderson,matthias2}. He passed away in 1980, right before the field of unconventional superconductivity would take off, but his legacy lives on as the reader will
see in many articles in this Special Issue.

One of Matthias' closest collaborators during that early period was Ted Geballe, one of the authors of this Special Issue. Ted has 
contributed longer than anybody else to the knowledge in this field over a distinguished 65 year scientific
career, making numerous seminal contributions to the discovery and understanding of many superconducting materials
covered in this volume. The editors would like to dedicate this Special Issue  to Ted on the year of his 95th birthday.

In the remaining part of this introductory article we provide a  brief overview of the contents of this Special Issue.
In the {\hyperlink{table_link} {Table}}, we list the classes of materials, the year of discovery, the highest $T_c$ in the class,
some physical properties, and whether it is believed to be conventional, possibly unconventional, or unconventional. 
A timeline  {\hyperlink{figure_link} {Figure}} summarizes progress to date.

\section{overview}

The article following this Introduction,
\href{http://www.sciencedirect.com/science/article/pii/S0921453415000362}{``What $T_c$ tells"}\cite{geballe}, by Geballe, Hammond and Wu, proposes that the value of $T_c$ itself as well as the response of
$T_c$ to various parameter changes such as ionic mass, composition, pressure or structure can give valuable clues on the
superconductivity mechanism. It discusses selected examples of this idea
 for both conventional and unconventional materials and
in so doing  gives a nice overview of the historical development of the field
of superconducting materials. The authors also discuss negative $U$ centers as a
possible non-conventional pairing  mechanism relevant to superconductivity of certain materials. 

The ensuing 32 articles cover each a different materials class, written by distinguished experimentalists in
each class. In the following we briefly discuss the 32 classes of materials
with 12, 9 and 11 classes in the C (conventional), P (possibly unconventional) and U (unconventional) categories, respectively.
By clicking on the class number  (e.g. {\bf C1}) the reader will be directed to the {\hyperlink{table_link} {Table}}, where clicking on the material class name
will bring the reader to the actual paper discussing that class. 

 In the closing section of this Special Issue, Greg Stewart \cite{stewart}, a former Ph.D. student of Ted Geballe, provides us with some
{ \href{http://www.sciencedirect.com/science/article/pii/S092145341500060X}{highlights of Ted's illustrious career}}. Finally, we have included an {\href{http://arxiv.org/abs/1504.02488}{Epilogue} \cite{epilogue}, where several major experimental contributors to the field of superconducting materials
have shared their views on past accomplishments and future hopes in this field.

\subsection*{ Conventional superconductors    }

\hypertarget{C1link} {}
{\noindent \bf \hyperlink{table_linka}{C1}:}
The first article in this category, ``Superconductivity in the elements,
alloys and simple compounds'' by Webb et al.\cite{c1}, gives a review of the earliest superconducting materials discovered, that were
known when BCS theory was proposed, and describes the extensions of BCS theory to include the retarded nature of the
phonon-induced effective electron-electron interaction, necessary for the understanding of deviations of the properties
of these materials from the predictions of simple BCS theory. It also recounts the  successes and failures of
theoretical efforts to explain the observed
$T_c$'s of elements and simple compounds using this theoretical framework.
\vs


\hypertarget{C2link} {}
{\noindent \bf \hyperlink{table_linka}{C2}:}
The second article by Stewart  \cite{c2} reviews the A15 compounds, discovered in 1954. The A15's are distinguished by the fact that for over 30 
years they were the highest $T_c$ materials known, and they were and are today the superconductors that are used in many high magnetic field applications.
They are believed to be almost prototypical Eliashberg electron-phonon driven superconductors, except that phonon anomalies seem to add some
interesting wrinkles, and peaks in the electronic density of states at the Fermi level play an important role in the quantitative understanding of their properties.
Stewart makes a number of interesting comparisons with other superconducting families, not the least of which is that Cs$_3$C$_{60}$, a doped fullerene that becomes superconducting at 38 K only under pressure, actually adopts an A15 lattice structure. This also makes it a member of the
growing family of materials that are insulating at ambient pressure but superconducting under pressure.
\vs

\hypertarget{C3link} {}
{\noindent \bf \hyperlink{table_linka}{C3}:}
The third article by Bustarrett \cite{c3}reviews doped semiconductors, a class discovered in the 60's that underwent a revival of interest starting in the mid-90's when higher $T_c$ materials were found. The carrier concentration in these materials is very low; for the most part, they are understood within the conventional framework (hence we included them in the first category), but the author notes that there are some puzzles such as Tl-doped PbTe that may require a different mechanism, as also discussed by Geballe et al. in their article.
\vs
 
\hypertarget{C4link} {}
{\noindent \bf \hyperlink{table_linka}{C4}:}
The 31 known superconducting elements at ambient pressure are metals.   In 1964 the first  non-superconducting 
element to become superconducting under pressure was discovered,  $Te$, a semiconductor at ambient pressure. 
Since then, many other semiconducting and insulating elements have been found to 
become metallic and superconducting at high pressures, as reviewed by Shimizu \cite{c4} in the
fourth article. The highest $T_c$ among insulating elements under pressure is sulfur with $T_c=17K$. The superconductivity in
this class is understood to arise from the conventional electron-phonon mechanism.
\vs

\hypertarget{C5link} {}
{\noindent \bf \hyperlink{table_linka}{C5}:}
Superconductivity in graphite intercalation compounds is reviewed in the fifth article, by Smith et al. \cite{c5} The first material in this class was 
discovered in 1965. The authors review the early history of
these materials, which have $T_c$'s of a few K, and the recent revival of interest with the discovery of superconductivity
in $C_6Ca$ and $C_6Yb$, with $T_c$'s up to $12K$.  They discuss the important role of
dimensionality and charge transfer,  the difficulties in understanding the different role for the superconductivity
of the intercalant metal band versus
the graphite $\pi$ and $\pi^*$ bands arising from $C$ $p_z$ orbitals, and the conflicting information from the large $Ca$ isotope 
shift observed in $C_6Ca$. The authors state that the pairing mechanism has always been an open question. Nevertheless, we
included this class in the first category because theoretical work on these materials has  focused on the
conventional mechanism.
\vs
\begin{table*}
 \hypertarget{table_link} {}
  \hypertarget{table_linka} {}
 \caption{Classes of superconducting materials. C (conventional), P (possibly unconventional) and
 U (unconventional).  The entries in the `Material class' column link to the paper on each class.
 The `Year' indicates which year the first material in the class was discovered.
The `Max $T_c$'  refers to ambient pressure except for C4 and C6. For `mag?', y/n indicates whether or not there are  magnetic
phases nearby in the phase diagram. `dim'=dimensionality of the structural part of the material believed to drive
superconductivity. `symm'=symmetry of the order parameter. Typical values of 
coherence length $\xi$, penetration depth $\lambda_L$ and gap ratio are given.  $dT_c/dP$ indicates the sign
of the change of $T_c$ with pressure for most materials in the class. }
 \begin{tabular}{l | c | c | c | c | c  | c | c  | c | c |c | c |c}
  \hline
&   Material  class & Year& Max $T_c$ &$T_c^{max}$ &   $\xi$   &   $\lambda_L$  & $2\Delta/$& $dT_c/$ &  mag? & dim& symm & Cate- \cr
  &  &   & material& (K)     &  ($\AA$)  & ($\AA$) &$k_BT_c$ &$dP$ &    & & & gory  \cr
 \hline  \hline
{\bf     \hyperlink{C1link} {C1} } & \href{http://www.sciencedirect.com/science/article/pii/S0921453415000647}{Elements},    & 1911 & $Nb$ & $9.5$ &   380 & 390 & 3.80 & +/- &  n & $3$&   $s$  &conv\cr
& \href{http://www.sciencedirect.com/science/article/pii/S0921453415000647}{alloys and simple compounds} & 1912 & $NbN$ & $17$ &   50 & 2000 &4.1  & +/- &  n & $3$&  $s$  &conv\cr
\hline
{\bf  \hyperlink{C2link}{C2}} &     \href{http://www.sciencedirect.com/science/article/pii/S0921453415000404}{A15's} & 1954  & $Nb_3Ge$ & $23.2$ &  55  &1000   & 4.2 &  + &n &$3$& $s$ &conv\cr
\hline
{\bf \hyperlink{C3link} {C3}}  & \href{http://www.sciencedirect.com/science/article/pii/S0921453415000489}{Doped semiconductors}& 1964 & $CB_x$  &10&  950 & 720 & 3.5& - &n & $3$  & $s$ &conv \cr \hline
{\bf   \hyperlink{C4link} {C4}} &  \href{http://www.sciencedirect.com/science/article/pii/S0921453415000623}{Insul. elements under pressure} &1964  &  S & 17 &     &   & & +&n & $3$&$s$  &conv \cr \hline
{\bf  \hyperlink{C5link} {C5}}&   \href{http://www.sciencedirect.com/science/article/pii/S0921453415000568}{Intercalated graphite}& 1965 & $C_6Ca$  & 11.5 &  380 & 720 & 3.6& + &n &$2$ &$s$  &conv \cr \hline
{\bf  \hyperlink{C6link} {C6}} &   \href{http://www.sciencedirect.com/science/article/pii/S0921453415000593}{Metallic elements under pressure} & 1968 &Ca   & 25 &     &  & &+/-&n & $3$&$s$  &conv \cr \hline
{\bf  \hyperlink{C7link} {C7}} & \href{http://www.sciencedirect.com/science/article/pii/S0921453415000441} {Hydrogen-rich materials}& 1970 & $PdD $ & $10.7   $& 400 &     & 3.8 & +/-& n & $3$&$s$  &conv \cr \hline
{\bf  \hyperlink{C8link} {C8}} & \href{http://www.sciencedirect.com/science/article/pii/S0921453415000507}{Layered t. m.  dichalcogenides}&1970  & $Nb S_2$  & 7.2 &  100  & 1250 &3.7  & - & n & $2$& $s$ &conv.\cr \hline
{\bf   \hyperlink{C9link}{C9}} &  \href{http://www.sciencedirect.com/science/article/pii/S0921453415000465} {Chevrel phases}& 1971 &   $PbMo_6S_8$  & $15$ &   30 &  3000& 4.7 & +/-  &y &  $3$ &  $s$&conv \cr \hline
{\bf  \hyperlink{C10link} {C10}} & \href{http://www.sciencedirect.com/science/article/pii/S0921453415000908} {Magnetic superconductors} &1972  & $ErRh_4B_4$  & 8.7  & 180   & 830 &4  & +/-  & y & $3$ & $s$ &conv \cr \hline
{\bf  \hyperlink{C11link}{C11}}  &  \href{http://www.sciencedirect.com/science/article/pii/S0921453415000143}{Thin films} &1978  &   &  &    & & &  &n & $2$&$s$  &conv\cr \hline
{\bf \hyperlink{C12link}{C12}}&   \hypertarget{table_linkb} {}  \href{http://www.sciencedirect.com/science/article/pii/S0921453415000519}{Magnesium diboride}&  2001& $MgB_2$  &  $39$ & 52   & 1400  & 4.5 & - &n & $2$ & $s$&conv\cr
 \hline
\hline   
{\bf  \hyperlink{P1link}{P1}} &  \href{http://www.sciencedirect.com/science/article/pii/S0921453415000398}{Bismuthates} & 1975 & $Ba_{1-x}K_xBiO_3$ & $34$ & 50& 5500& 4& - &n & $3$ & s &poss unc \cr\hline
{\bf  \hyperlink{P2link}{P2}}&  \href{http://www.sciencedirect.com/science/article/pii/S0921453415000416} {Fullerenes} &  1991 & $RbCs_2C_{60}$  & $33$ & 30  & 4500  &3.5-5.0  & - &n &$0$ & s &poss unc\cr \hline
{\bf  \hyperlink{P3link} {P3}}& \href{http://www.sciencedirect.com/science/article/pii/S0921453415000611}{Borocarbides} &  1993 &  $YPd_5B_3C_{0.3}$  &23& 100   & 1000   & 4 & +/- & y ,n& $2$ & s+g? &poss unc \cr \hline
{\bf \hyperlink{P4link} {P4}}&    \href{http://www.sciencedirect.com/science/article/pii/S0921453415000581}{Plutonium compounds} &   2002  & $PuCoGa_5$  & 18.5& 16 & 2400&5-8& +/- &y & $2$ &$d$ &poss unc \cr \hline
{\bf  \hyperlink{P5link}{P5}} &  \href{http://www.sciencedirect.com/science/article/pii/S0921453415000556}{Interface superconductivity}&  2007   & LaAlO$_3$/SrTiO$_3$ & .35  &600  &  & &  & y &$2$ & $ $ &poss unc \cr \hline
{\bf   \hyperlink{P6link} {P6}} &\href{http://www.sciencedirect.com/science/article/pii/S0921453415000428}{Aromatic hydrocarbons}&  2010   &K-doped DBP & 33 & 180 & 770 &  & +/- &n &  $3$ &  &poss unc \cr \hline
{\bf  \hyperlink{P7link}  {P7}} & \href{http://www.sciencedirect.com/science/article/pii/S0921453415000453}{Doped top. ins.} & 2010    &Cu$_x$(PbSe)$_5$(Bi$_2$Se$_3$)$_6$  & 3 & 110 & 13000 &  &  & n & $2$&  &poss unc\cr \hline
{\bf  \hyperlink{P8link}{P8}} &  \href{http://www.sciencedirect.com/science/article/pii/S0921453415000520}{$BiS_2-$based materials}& 2012 &$YbO_{0.5}F_{0.5}BiS_2$ &5.4& 53 & 5000 & 7.2& +/- &n & $2$ & s &poss unc \cr \hline
{\bf \hyperlink{P9link}{P9}} &  \href{http://www.sciencedirect.com/science/article/pii/S0921453415000544} {Unstable/elusive sc}&1946  & $C-S$   &300?  &  &   &   & & n &$2$&  &poss unc \cr \hline
\hline
{\bf  \hyperlink{U1link}{U1}}&   \href{http://www.sciencedirect.com/science/article/pii/S0921453415000714}{Heavy fermions}& 1979 & $UPd_2Al_3$    &2 & 50 &4000& &+/- & y & $3$ & $d, p$ &unconv \cr \hline
{\bf  \hyperlink{U2link} {U2}} & \href{http://www.sciencedirect.com/science/article/pii/S092145341500057X} {Organic  charge-transfer} & 1980 &   $(BEDT-TTF)_2X$  &13.4 & 100 &  5000 &4.4 & - & y &  $1$, $2$&$d$  &unconv\cr \hline
{\bf  \hyperlink{U3link}  {U3}} & \href{http://www.sciencedirect.com/science/article/pii/S0921453415000878}{Cuprates hole-doped}  & 1986 & $HgBa_2Ca_2Cu_3O_9$& $134$ & 20 & 1200  &4.3 & +&y & $2$ & $d$& unconv \cr \hline
{\bf  \hyperlink{U4link} {U4}}&  \href{http://www.sciencedirect.com/science/article/pii/S0921453415000635}{Cuprates e-doped}  & 1989 & Sr$_0.9$La$_x$CuO$_2$  & $40$  & 50  &2500   &3.5 & - &y &$2$&$d$  &unconv \cr \hline
{\bf  \hyperlink{U5link} {U5}} & \href{http://www.sciencedirect.com/science/article/pii/S0921453415000660}{Strontium ruthenate} & 1994 & $Sr_2RuO_4$     &1.5 & 660 & 1500  &  & - & y & $2$ &  $p$&unconv\cr \hline
{\bf  \hyperlink{U6link}  {U6}} &  \href{http://www.sciencedirect.com/science/article/pii/S0921453415000490} {Layered nitrides}& 1996  & $Ca(THF)HfNCl$&26 & 60& 4700  & 2.9-10& - &n & $2$ & $d+id$ &unconv \cr \hline
{\bf  \hyperlink{U7link} {U7}} & \href{http://www.sciencedirect.com/science/article/pii/S0921453415000532} {Ferromagnetic sc} &2000& UGe$_2$   & 0.8 & 100   & $\sim 10^4$   &     & +/-   & y & $3$& p   &unconv \cr \hline
{\bf   \hyperlink{U8link} {U8}}&  \href{http://www.sciencedirect.com/science/article/pii/S0921453415000374}{Cobalt oxyde hydrate} & 2003&  Na$_x$(H$_3$O)$_z$CoO$_2 \cdot y$H$_2$0 &4.7 & 100 &7000   &4.3-4.6  & - &y &  $2$ & ?  &unconv \cr \hline
{\bf \hyperlink{U9link}  {U9}} &  \href{http://www.sciencedirect.com/science/article/pii/S092145341500043X}{Non-centro-symmetric} & 2004    &  SrPtSi$_3$  & 2 & 60    & 8000    & &   & y  & $3$&s/p   &unconv \cr \hline
{\bf   \hyperlink{U10link}{U10}} & \href{http://www.sciencedirect.com/science/article/pii/S0921453415000477}{Iron pnictides} & 2008 & $SmFeAsO_{0.85}$  & $55$ & 10-50 &2000  &7.5 & +/-&y &  $2$& $s \pm$& unconv \cr \hline
{\bf \hyperlink{U11link} {U11}} & \href{http://www.sciencedirect.com/science/article/pii/S0921453415000386} {Iron chalcogenides} & 2008 &  $Na_xFe_2Se_2$   &46 & 20  & 2000  & 3.8 &+&y &  $2$ & $s$&unconv\cr \hline
 \hline
 \end{tabular}
\end{table*}

\hypertarget{C6link} {}
{\noindent \bf \hyperlink{table_linka}{C6}:}
The first non-superconducting metal found to become superconducting under pressure was Ce, in 1968. Since then,
many more non-superconducting metals at ambient pressure were found to become superconductors under high pressure, some with
remarkably high $T_c$, as reviewed by Hamlin \cite{c6}. 
Hamlin discusses the various ways that this has been understood within the conventional framework, namely,
an increased $s$ to $d$ electron transfer, phonon softening, or suppression of spin fluctuations. He also points out a remarkable negative isotope effect observed in $Li$ at high pressure that
may derive either from an unconventional superconductivity mechanism or anharmonicity in the phonon spectra. 
He remarks that conventional theory
has difficulty differentiating between elements that display a high $T_c$ at high pressures and others with low or zero
$T_c$ under high pressure. 
\vs

 \hypertarget{C7link} {}
 {\noindent \bf \hyperlink{table_linka}{C7}:}
As discussed by Struzhkin \cite{c7}, Eliashberg theory predicts
that phonon frequencies contributing mostly to $T_c$ are of order $10\times k_BT_c/\hbar$, thus favoring high vibrational frequencies to
reach  high $T_c$'s within the conventional mechanism. Hydrogen-rich materials  are expected to
have the highest vibrational frequencies and hence have been investigated extensively. Struzhkin reviews the early work on
metal hydrides commencing in 1970 and the finding of a negative hydrogen isotope effect, that first raised doubts on the validity of the electron-phonon
mechanism for superconductivity in these materials, and then was understood as arising from anharmonicity of hydrogen vibrations.
He then reviews the more recent intensive efforts searching for superconductivity in hydrogen-rich materials under high pressures, 
the reported finding of $T_c\sim20K$ in compressed $SiH_4$, the recent indications of $T_c\sim 190K$ in compressed $SH_2$, and
the theoretical predictions of high $T_c$ in compressed polyhydrides of alkali and alkaline earth metals, in particular 
$T_c$ as high as $235K$ predicted for $CaH_6$.
\vs
 
\hypertarget{C8link} {}
{\noindent \bf \hyperlink{table_linka}{C8}:}
Intercalated and pristine layered transition metal dichalcogenides are 
quasi-2-dimensional materials that often exhibit competing charge-density-wave instabilities. Klemm \cite{c8} reviews this class of materials and
points out that many of their 
properties are strikingly similar to properties seen in the cuprates and iron pnictides, such as 
pseudo gap behavior and incoherent c-axis transport, and in that sense should be regarded as `unconventional'. However, since
there has been no suggestion that the pairing mechanism is anything other than the electron-phonon interaction, we have
included this class among the conventional superconductors. These materials show no traces of magnetism.
\vs

\hypertarget{C9link} {}
{\noindent \bf \hyperlink{table_linka}{C9}:}
Pe\~na \cite{c9} discusses Chevrel phases, a rich  class of materials discovered in 1971 with $T_c$'s up to $15K$ and very large upper
critical fields.  Some of these compounds containing rare earth elements exhibit coexistence of superconductivity and magnetic order and an exchange field compensation effect leading to magnetic field-induced superconductivity (Jaccarino-Peter effect).  The author classifies these superconductors  as ``exotic''
given that the ratio $T_c/T_F$ ($T_F$=Fermi energy) falls between $1/100$ and $1/1000$, in contrast to more conventional materials
where $T_c/T_F<1/1000$. Even though there have been a few suggestions of unconventional order parameter symmetry and mechanism, by and large they are regarded as conventional. Hence we have included them in the first category.
\vs

\hypertarget{C10link} {}
{\noindent \bf \hyperlink{table_linka}{C10}:}
The interplay between superconductivity and magnetism in conventional superconductors is reviewed by Wolowiec et al. \cite{c10}.  The authors describe some of the extraordinary phenomena that are found in conventional superconductors containing ions with partially-filled d- or f-electron shells that carry magnetic moments in both paramagnetic and magnetically-ordered states.  Two general cases are considered, one in which the ions that carry the magnetic moments are dissolved in a superconducting host as impurities, and another in which they occupy an ordered sublattice in a superconducting compound.  In both cases, the conventional superconductivity is associated with two sets of electrons, an itinerate set of electrons that are involved in the superconductivity and a localized set of d- or f-electrons that carry magnetic moments, that interact with one another via the exchange interaction.  Some remarkable phenomena are observed such as reentrant superconductivity due to the Kondo effect or ferromagnetic order, coexistence of superconductivity and antiferromagnetic order, and magnetic field-induced superconductivity, etc. 
This case also serves as a background for phenomena encountered in materials in which the superconductivity and magnetism involve the same set of electrons, such as heavy fermion, cuprate, and iron pnictide and chalcogenide superconductors, where the unconventional superconductivity is often found in a ``dome-shaped" region near the solute composition or pressure where an antiferromagnetic transition is suppressed towards 0 K.  
\vs
 
\hypertarget{C11link} {}
{\noindent \bf \hyperlink{table_linka}{C11}:}
Superconductivity in thin films has been a focus of study across different classes of materials that are superconducting.
Thin films have been useful for applications; in addition, their study allows us to understand the effects of
dimensionality on the superconducting phase transition. Lin, Nelson and Goldman \cite{c11} focus on the superconducting
insulator transition as a paradigm for a quantum phase transition, particularly as the two dimensional limit is approached.
\vs

\hypertarget{C12link} {}
{\noindent \bf \hyperlink{table_linka}{C12}:}
The final class in this category contains a single compound, $MgB_2$, discovered in 2001 by Akimitsu and coworkers, with $T_c=39K$, which is remarkably high
for a simple binary compound with only s- and p-electrons. For this reason, initially various alternative explanations of its superconductivity
were proposed.  Bud'ko and Canfield \cite{c12} provide a comprehensive overview of various properties of this
compound, from synthesis to basic transport and thermodynamic properties to mechanism. They present the prevailing view
that $MgB_2$ is a conventional electron-phonon driven superconductor, albeit with two well-defined superconducting bands and two gaps. This view is supported by
observation of a partial isotope effect and by the  agreement of
conventional Eliashberg calculations with the measured $T_c$.
Bud'ko and Canfield also note that, while the discovery of $MgB_2$ has not been followed by a family of related superconductors, a considerable
effort has been made to develop this material for MRI magnets and other applications. 
\vs

\subsection*{Conventional or unconventional superconductors?}

In the second category, ``possibly unconventional superconductors'' (P), we have grouped materials that for one reason or another show indications that their
pairing mechanism may not be the electron-phonon interaction that drives superconductivity according to  the conventional theory of superconductivity.
\vs
\hypertarget{P1link} {}
{\noindent \bf \hyperlink{table_linka}{P1}:}
When $BaPb_{1-x}Bi_xO_3$ was discovered in 1975 \cite{sleight} it was pointed out that its $T_c$ was exceptionally high for an oxide and
higher than that of any superconductor not containing a transition metal known at that time. Sleight \cite{p1} discusses this class of materials, the bismuthates, emphasizing their chemistry. These low-carrier-concentration oxides are often said to be precursors to the high $T_c$ cuprates, having a perovskite structure with $BiO_2$ instead of $CuO_2$. Unlike the cuprates they are three-dimensional materials and display no magnetism; instead, like the 
transition metal dichalcogenides (C8), they often exhibit charge-density waves. $BaPb_{1-x}Bi_xO_3$ has $T_c=13K$,  while the highest $T_c$ in this class is a 
remarkable $34K$ for
$Ba_{1-x}K_xBiO_3$. Sleight emphasizes the tendency of the $Bi$ ion to disproportionation
($Bi^{4+}\rightarrow Bi^{3+}+Bi^{5+}$)  and the resulting possible negative $U$ mechanism,  also discussed in the article by Geballe et al., leading to what is often called `bipolaronic' superconductivity. Many proposed explanations of the superconductivity in these materials have involved non-conventional mechanisms,  while
others use the  conventional BCS-Eliashberg electron-phonon theory.
\vs

\hypertarget{P2link} {}
{\noindent \bf \hyperlink{table_linka}{P2}:}
The remarkable discovery of the $C_{60}$ molecule was followed only a few years later by the discovery of superconductivity in
alkali-doped $C_{60}$ with critical temperatures as high as $33K$. This and $BaKBiO$  would have set records for
high $T_c$, had the cuprates not been discovered a few years earlier. Ramirez \cite{p2}  provides a summary of experimental results that all seem to indicate a somewhat conventional electron-phonon
mechanism for the superconductivity, although he notes that some open questions remain, particularly in the context of insulating phases, with 
evenly-doped compounds bracketing the oddly-doped superconducting compounds.
\vs

\hypertarget{P3link} {}
{\noindent \bf \hyperlink{table_linka}{P3}:}
Quaternary borocarbides, $YNi_2B_2C$, being the prototypical one, are reviewed by Mazumdar and Nagarajan \cite{p3}. There are
about 60 variations, by substitution of $Ni$ by $Pd, Cu, Co$ or $Pt$ and $Y$ by various rare earths, 25 of which are 
superconducting, with the highest $T_c\sim 23K$. They exhibit a rich variety of phenomena including valence fluctuations, heavy fermion
behavior, and coexistence of superconductivity and magnetic order. There are indications of unconventional
superconductivity, such as an anisotropic gap, as well as evidence of conventional superconductivity, such as a $B$ isotope effect, 
and it is regarded as an open question whether these materials are unconventional or conventional superconductors.
\vs

\hypertarget{P4link} {}
{\noindent \bf \hyperlink{table_linka}{P4}:}
Plutonium based compounds, a 4-member class  of which $PuCoGa_5$ has
the highest $T_c=18.5K$, are reviewed by Sarrao et al. \cite{p4}, who suggest that
they are a bridge between heavy fermion superconductors (with much lower $T_c$'s)
and high $T_c$ cuprates. They are related
to  $Ce-$ and $U-$ based heavy fermion superconductors;
e.g., $PuCoGa_5$ can be viewed as layers of $PuGa_3$ and $CoGa_2$, similarly to $CeMIn_5$ that can be regarded as layers of $CeIn_3$ and $MIn_2$.
They exhibit heavy fermion behavior, valence fluctuations,
and a quasi-2D Fermi surface.  There is evidence for a $d-$wave gap from point-contact tunneling and
power-law behavior in spin-lattice relaxation rate and penetration depth,
as well as an absence of a Hebel-Slichter peak in NMR. While the
evidence for unconventional superconductivity is regarded as non-conclusive
and phonon-mediated pairing has also been proposed, the authors 
suggest that unconventional superconductivity in these materials may be
due to spin-fluctuation mediated pairing or induced by valence fluctuations in a
strongly correlated mixed valent normal state. They also remark that one should not assume that
all the $Pu-115$ superconductors have the same pairing mechanism.
An interesting peculiarity of these materials is that the superconductivity gets
degraded with time due to radiation damage from radioactive Pu.
\vs

\hypertarget{P5link} {}
{\noindent \bf \hyperlink{table_linka}{P5}:}
Interface superconductivity is in some ways a natural extension of the ultra-thin film superconductivity discussed in C11.
Gariglio et al. \cite{p5} focus on the two-dimensional superconductivity at the interface between the two band insulators
LaAlO$_3$ and SrTiO$_3$. The ability to exercise careful control of the properties of this interface through the application
of electric fields is also highlighted. While the superconducting $T_c$'s have remained quite low, these systems provide
a plethora of physics, some of which is connected, for example, to the cuprates (quasi 2D and pseudo gap behavior), while
other behavior (significant spin-orbit coupling) shows similarities with the doped topological superconductor family of materials, and
may cause non-BCS behavior in the superconducting properties of these interfaces.  
\vs

\hypertarget{P6link} {}
{\noindent \bf \hyperlink{table_linka}{P6}:}
Superconductivity in aromatic molecules was suggested a long time ago by
Fritz London. In 2010, it was finally found in alkali doped aromatic hydrocarbons, 
as reviewed by  Kubozono and coworkers \cite{p6}, for example,
$K_x$picene (five benzene rings) with $T_c\sim 18K$. The authors point out that for this class of materials, the shielding fraction is usually very low and the highest priority research is to increase it.
For $K_x$picene, they suggest that superconductivity may be
explainable by the conventional BCS phonon-mediated mechanism
aided by a high density of states,
high frequency intramolecular phonon modes, and contribution from intermolecular
phonon modes. 
However, superconductivity was later found in 
K-doped dibenzopentacene with $T_c=33K$. Depending on the compound,
$T_c$ decreases or increases with pressure, the latter of which cannot be easily explained within the BCS framework.
The authors note a correlation between $T_c$ and the number of benzene rings.
\vs

\hypertarget{P7link} {}
{\noindent \bf \hyperlink{table_linka}{P7}:}
Sasaki and Mizushima \cite{p7} review various ideas surrounding superconducting doped topological materials.
Less is known about these materials --- they are primarily motivated by the search for specific exotic properties, such as
so-called Majorana fermions. The authors describe a number of candidate materials that are currently thought to be
examples of superconducting doped topological materials, and they discuss a series of smoking-gun experiments to
look for topological superconductors.
\vs

\hypertarget{P8link} {}
{\noindent \bf \hyperlink{table_linka}{P8}:}
Superconductivity in the recently discovered class of layered
$BiS_2$-based compounds is reviewed by Yazici et al. \cite{p8}.
These materials are interesting because the structure of the $BiS_2$ layers is
similar to the $CuO_2$ layers in cuprates and $FeAs$ layers in iron pnictides,
but there is no magnetism in many of the materials in this class. 
The maximum $T_c$ is  $5.4K$ at ambient
pressure  and  $\sim 10K$ under pressure. 
Conventional electron-phonon pairing has been suggested, as well as 
various other scenarios such as 
Van Hove singularities, Fermi surface nesting
and spin fluctuations.
The authors remark that there is no consensus on
the origin of superconductivity in these materials.  There is some evidence for two superconducting
gaps, the superfluid density is low, and the gap ratio is very high.
\vs

\hypertarget{P9link} {}
{\noindent \bf \hyperlink{table_linka}{P9}:}
To conclude this category, unstable and elusive superconductors are
reviewed by Kopelevich et al. \cite{p9}. These materials show signals of
superconductivity, sometimes at very high temperatures, but bulk superconductivity has not been conclusively established. The authors review early (1978) 
results on $CuCl$
showing diamagnetic signals below $170K$ together with a drop in resistivity,
with these phenomena occurring up to $T\sim240K$ in films.
More recently (1999), there have been indications of superconductivity around $90K$ in
$K$-doped $WO_3$, and recent indications of transient superconductivity
in  sulfur-doped graphite or amorphous carbon at various high
temperatures, including room temperature. The origin of these phenomena is
unknown, so we have grouped this class in the possibly
unconventional category.

\subsection*{Unconventional superconductors}
For the materials in this  category  (U) there is a consensus that the superconductivity does not originate in the
conventional BCS-Eliashberg electron-phonon mechanism.
\vs

\hypertarget{U1link} {}
{\noindent \bf \hyperlink{table_linkb}{U1}:}
The first superconductors clearly identified as unconventional   were
heavy fermion superconductors such as $CeCu_2Si_2$ (the first in this
class, discovered in 1979),   $UBe_{13}$, $UPt_3$
and $URu_2Si_2$. 
These materials, reviewed by White et al. \cite{u1}, have enormous densities of states and effective masses for the 
conduction electrons, enhanced by a factor as high as $\sim1000$  with respect to that of
a conventional metal. White et al.  point out that these materials cannot be
conventional superconductors because the ordering of energy 
scales, $T_F<\theta_D$, with $T_F$ the effective Fermi temperature for the
heavy 	quasiparticles and $\theta_D$ the Debye temperature, is opposite of that of conventional superconductors where $\theta_D<<T_F$. The large specific heat jump at $T_c$ shows
that Cooper pairs are formed from heavy quasiparticles and the gap $\Delta(\vec{k})$ 
exhibits line or
point nodes at the Fermi surface, unlike a conventional s-wave gap. The heavy fermion behavior and superconductivity exhibited by these materials usually involve $f$-electrons from $Ce$, $U$, $Yb$, $Pr$, etc (although there are some recent reports
of $d$-electron heavy fermion systems), electron bands are very narrow, and the electrons are strongly correlated.  Many heavy fermion compounds have complex temperature versus solute composition or pressure phase diagrams in which superconductivity appears in the proximity of, and sometimes coexists with, antiferromagnetic order.  The superconducting $T_c$'s, however, are disappointingly low, in the range  of $1K$, except for the small subclass of $Pu$ materials treated here as a separate class and covered by Sarrao et al. as discussed earlier  (P4).
\vs

\hypertarget{U2link} {}
{\noindent \bf \hyperlink{table_linkb}{U2}:}
Superconducting organic charge transfer compounds were discovered in 1980,
culminating a long search for `excitonic' superconductivity in low-dimensional 
organic materials,
as reviewed by Brown \cite{u2}. Organic molecules stack in quasi-1D or quasi-2D arrangements,
and there are competing charge and spin density wave phases nearby. Correlations
play a significant role, and it has been proposed that either charge fluctuations or
antiferromagnetic  spin fluctuations mediate the pairing, mostly the latter.  The NMR relaxation rate shows no Hebel-Slichter peak, and power law behavior 
for various properties at low temperatures is seen, similar
to the cuprates. From this and other evidence it is inferred that the gap function has nodal structure, consistent with $d-$wave symmetry.
The highest $T_c$ is $\sim 13K$.
\vs
 
 \hypertarget{U3link} {}
 {\noindent \bf \hyperlink{table_linkb}{U3}:}
As Chu and coauthors \cite{u3} remark, hole-doped cuprates, discovered by Bednorz and Muller in 1986, 
``have ushered in the modern era of high temperature superconductivity''. This vast class of layered copper oxide compounds
with perovskite-like structures comprises more than 200 superconductors belonging to 7 families, as surveyed by Chu et al., with $T_c$'s up to
$134K$ at ambient pressure, and is the only class of materials with confirmed critical temperatures above 
 liquid nitrogen temperature (77K).
The essential structural feature of these materials is the $CuO_2$ planar lattice. 
Superconductivity occurs over a limited range of hole-doping a parent antiferromagnetic insulator, with a ``dome-shaped'' curve of
$T_c$ versus carrier concentration. The most anomalous types of behavior are found in the underdoped region.
Many workers have focused on understanding the anomalous normal state properties of these materials
such as the linear temperature dependence of the resistivity and pseudogap behavior.
Non-conventional behavior in the superconducting state includes a large positive pressure dependence of $T_c$,
absence of an isotope effect in the optimally doped and overdoped regimes, power-law low temperature behavior of
various quantities, and nodal structure of the gap inferred from photoemission and phase sensitive experiments.
The superconducting gap is believed to have d-wave symmetry and most workers believe pairing is 
induced by spin fluctuations.
\vs

\hypertarget{U4link} {}
{\noindent \bf \hyperlink{table_linkb}{U4}:}
Fournier \cite{u4} reviews the younger 'siblings' of the hole-doped cuprates, i.e. the electron-doped cuprates.
This family also has CuO$_2$ planes, but, unlike the hole-doped materials, there are no apical oxygen atoms.
The author emphasizes that two structures give rise to this distinct characteristic, the $T^\prime$ and the infinite
layer phase. The latter subgroup has family members with $T_c$ as high as 40 K. Common elements to both hole-doped
and electron-doped cuprates are that $T_c$ also has a dome-shaped dependence
as a function of carrier concentration, normal state properties are anomalous, and the symmetry of the order parameter is
likely d-wave, although the author notes that this last point remains controversial. The author also remarks that transport and photoemission
data indicate that both electrons and hole carriers contribute to the transport in these materials.
\vs

\hypertarget{U5link} {}
{\noindent \bf \hyperlink{table_linkb}{U5}:}
Sr$_2$RuO$_4$, 
first synthesized in 1959 and found to be superconducting in 1994, has a layered perovskite crystalline structure like the high $T_c$
cuprates; however, its $T_c$ is only $1.5$ K.
Liu and Mao \cite{u5} provide an overview of superconducting properties; in particular they note the key experiments
that indicate a superconducting order parameter with $p$-wave symmetry, and other features in common with superfluid $^3$He. This material also has the intriguing property of having
superconductivity suppressed by hydrostatic pressure but enhanced by uniaxial pressure along the $c$-axis. The mechanism for
superconductivity remains unknown in this material.
\vs

\hypertarget{U6link} {}
{\noindent \bf \hyperlink{table_linkb}{U6}:}
Layered nitrides, discovered in 1996, are reviewed by Kasahara and coworkers \cite{u6}. These insulating compounds have 
layers of nitrogen and a metal ($Ti$, $Zr$, $Hf$) separated by halogen ($Cl$, $Br$, $I$) layers. Upon intercalation with alkali or
alkaline earth atoms, they become metallic and superconducting at a critical doping, and $T_c$ decreases upon further doping. 
The gap is isotropic for low doping and becomes anisotropic for higher doping level, the gap ratio decreases with doping, and the gap
goes from isotropic to anisotropic. There is no evidence for nodes in the gap.  The density of states is very low and the electron phonon interaction is not strong enough to account for the observed $T_c$'s. In addition, the electron-phonon interaction strength 
(measured by Raman) and the density of states increases with doping, yet $T_c$ decreases. These findings and the facts
that  the $N$ isotope coefficient is vanishingly small and there is  no Hebel-Slichter peak, indicate that the electron-phonon interaction is not
responsible for the superconductivity.
Pairing mechanisms that have been suggested for these materials include acoustic plasmons and charge or spin-fluctuation mediated
pairing, and the gap symmetry is suggested to be $d+id$. In contrast to other layered superconductors, the parent compounds are band
insulators with   no magnetic nor charge-density-wave order.
\vs

\hypertarget{U7link} {}
{\noindent \bf \hyperlink{table_linkb}{U7}:}
A material that is both ferromagnetic and superconducting was first discovered in 2000, and  now a handful are known to exist albeit at
low ($< 1 $K) temperatures. They remain of special interest for a variety of reasons --- these two phases shouldn't coexist in the first place, 
their coexistence could be exploited for spintronic-type applications, and their critical magnetic fields are extraordinarily high. Huxley \cite{u7}
gives a brief summary of their known properties. In the original superconducting material (UGe$_2$), pressure is required for superconductivity
to occur, but, as reviewed here, other materials in this family are superconducting at ambient pressure.
\vs

\hypertarget{U8link} {}
{\noindent \bf \hyperlink{table_linkb}{U8}:}
Sakurai {\it et al.} \cite{u8} provide an update on the more recently discovered cobalt oxide hydrates, the only known Co oxide superconductor, with $T_c = 4.7$ K.
These materials are two-dimensional, but separate themselves from other examples because
of their triangular lattice structure. The superconducting gap function has a line-nodal structure. They are strongly type-II, but significant controversy remains concerning other superconducting properties,
including the symmetry of the order parameter. A concise summary of some of the contradictory measurements is provided in this review.
\vs

\hypertarget{U9link} {}
{\noindent \bf \hyperlink{table_linkb}{U9}:}
Kneidinger et al. \cite{u9} present a review of superconductivity in non-centrosymmetric materials. They remark how many of the 
measured properties are consistent with BCS theory, although this does not necessarily imply the traditional electron-phonon mechanism, and
some of the materials also display non-BCS characteristics indicating unconventional superconductivity presumably arising from strong electronic correlations. 
They also note that this material family may ultimately be the source of materials for the topological superconductor class. 
A number of properties for various compounds in this class are presented, with references for more comprehensive lists and discussions.
\vs

\hypertarget{U10link} {}
{\noindent \bf \hyperlink{table_linkb}{U10}:}
In the class of iron pnictide superconductors discovered in 2008 and reviewed by Hosono and Kuroki \cite{u10},
superconductivity results from doping a parent antiferromagnetic metal or Pauli paramagnetic metal with either electrons or holes.
There are 7 types of parent materials each with a somewhat different structure, 
all containing $FeAs$ layers separated by spacer layers where dopants are introduced. 
The authors note that the same materials with either Co, Zn, Cr or Mn instead of Fe do not give rise to superconductors, while Ni 
gives rise to superconductors with much lower $T_c$, and hence that Fe must  play a central role in the superconductivity of these 
materials which reaches a high $T_c$ of 55K. Superconductivity emerges when antiferromagnetism disappears by carrier doping.
Electron-phonon interactions cannot account for the high $T_c$'s observed, hence both spin and orbital fluctuation mediated pairing 
mechanisms have been discussed. An $s\pm$ state has been proposed for those materials that have both electron and hole
Fermi surfaces, with the gap being of opposite sign in electron and hole parts, induced by spin fluctuations, and
alternatively a $s++$ gap with the same sign on the entire Fermi surface where pairing is induced by orbital fluctuations. 
\vs

\hypertarget{U11link} {}
{\noindent \bf \hyperlink{table_linkb}{U11}:}
Closely related to the Fe-pnictide superconductors are the Fe-chalcogenides, of which FeSe has the simplest crystal structure.
Chang et al. \cite{u11} describe a variety of  novel synthesis methods to prepare nanomaterials and thin films of these compounds. When intercalated
with alkali metals, or in thin film form including down to a monolayer, or under pressure, these materials attain remarkably
high $T_c$'s above $30K$. ARPES experiments show evidence for a nodeless isotropic gap and only electron pockets, hence  the $s+/-$ scenario proposed for some iron pnictides does not apply. The authors mention electronic nematicity, orbital and spin fluctuation pairing scenarios as
possibly relevant to   the iron based superconductors, leaning towards the orbital scenario for the Fe-chalcogenides.

\section{closing remarks}

The 32 classes of superconducting materials surveyed above and covered in detail in the   articles in this Special Issue present a complicated and confusing  picture. They exhibit a rich variety of other phenomena besides their superconductivity.
Presumably, some of  these phenomena are intimately related to the superconductivity while others are ``red herrings'' that are associated with the
other underlying interactions and excitations in the material but not directly related to the superconductivity. 
We don't know   which are which. 
Similarly, the superconductivity in the different classes has some features in common and others that appear different, but it is not always easy to know whether
the differences are extrinsic or intrinsic.

In thinking about which material class may spawn
the next higher temperature superconductor or even
a room-temperature superconductor, we realized
we have omitted some potential candidates. One  is ``transient superconductors'': there has been some recent 
 evidence that when hit with an infrared light pulse,  some cuprates can develop pairing correlations and perhaps superconductivity at room temperature,
  lasting for a few picoseconds  \cite{transient}. 
If so, for these materials the challenge will not be to raise the $T_c$ but to raise the lifetime (i.e., stabilize the
superconductivity).  Layer-by-layer deposition  of thin films  
by molecular beam epitaxy \cite{bozovic} potentially could be a route  to new superconducting materials that do not form by conventional methods. 
Also high pressure synthesis methods \cite{hps} could lead to new classes of  high $T_c$ superconductors. These would be  ways to
implement one of Matthias' recipes towards high $T_c$ materials \cite{matthias}: ``From now on, I shall look for systems that should
exist, but won't $-$ unless one can persuade them''.
Finally,   while the high pressure frontier is
amply covered in the articles by Hamlin, Shimizu, Struzhkin and other articles in this Special Issue, there is no article specifically covering insulating compounds
that become superconducting under pressure such as $CsI$\cite{csi}, $C_6I_6$ \cite{c6i6}, 
$C_6I_4O_2$ \cite{iodanil}. Given the many surprises already found in superconductivity under  pressure, 
the first room-temperature superconductor could well emerge from a simple insulating compound under ultrahigh pressure, not necessarily containing hydrogen which
is where the focus of interest has been in this respect \cite{struzhkin}.

 \begin{figure}
  \hypertarget{figure_link} {}
\resizebox{8.5cm}{!}{\includegraphics[width=7cm]{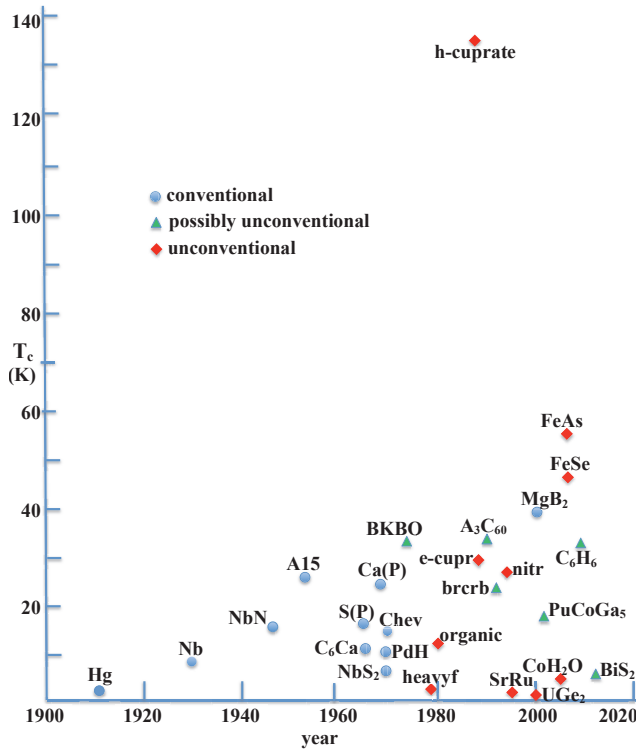}}
\caption {  Highest critical temperature of the different classes of superconducting materials
at ambient pressure (except for $Ca(P)$, $S(P)$), vs year that the class was discovered,
not the year the highest $T_c$ material in the class was found (except for $Nb$, $NbN$). 
$NbN$=simple compounds, heavyf=heavy fermions, brcrb=borocarbides, nitr=layered nitrides, $NbS_2$=intercalated transition metal dichalcogenides, Chev=Chevrel phases, $BKBO$=bismuthates, $SrRu$=Strontium Ruthenate,
$C_6Ca$=intercalated graphite,  $C_6H_6$=aromatic hydrocarbons, $CoH_2O$=cobalt oxide hydrides,  $UGe_2$=ferromagnetic superconductors, $PdH$=hydrogen-rich materials, 
$PuCoGa_5$=plutonium compounds, Ca(P)=metallic elements under  pressure, 
S(P)=insulating elements under  pressure,
e-cupr=electron-doped cuprates, h-cuprate=
hole-doped cuprates.}
\label{figure1}
\end{figure} 

 {\hyperlink{figure_link} {Fig. 1}} shows the highest critical temperature (at ambient pressure) within each class of materials covered  in this Special Issue versus the year when the class was  discovered. 
It is clear from the figure that there is little correlation between the magnitude of $T_c$ and whether the material
is conventional, unconventional or possibly unconventional.

How many different types of superconducting states exist in nature? How many different
types of pairing mechanisms are realized in materials? Does a pairing mechanism uniquely define a resulting superconducting state?
Does a single pairing mechanism act in each material or can more than one pairing mechanism coexist in a material? In the latter case,
do different pairing mechanisms enhance one another, or compete with one another, or contribute independently to raising $T_c$? These are some of the many unanswered questions.

 There are two natural straight lines one can draw in  {\hyperlink{figure_link} {Fig. 1}}. The first one connects Hg in 1911 to the hole-doped cuprates in 1986, the second connects
Hg to the iron pnictides (FeAs)  in 2008, leaving cuprates out as an anomaly. The conservative second line predicts that we will see room temperature
superconductors in the year 2464, the first line predicts that  this will happen in the year 2082. Neither is good enough. 
We need exponential rather than linear growth in $T_c$. For that,  the empirical-serendipitous approach that has brought us from $Hg$ in 1911 to where we are in 2015 needs to be supplemented with demonstrably effective  theoretical
guidelines to further  focus the search and reduce the amount of phase space being explored. Such   guidelines are currently lacking, and it can be plausibly argued that currently   researchers
can't see the ``forest for the trees''.
We hope that this Special Issue, by   juxtaposing  the specifics of each class of superconducting materials against the backdrop of all the other classes, will help 
to improve this situation. We are very  grateful to all the authors that contributed excellent articles to this Special Issue to help achieve this goal.


\begin{references}

  \bibitem{theories}See references in  W. L. Ginsburg, ``Der gegenw\"artige Stand der Theorie der Supraleitung'', Fortschritte der Physik, {\bf 1}, 101 (1953), and in
J. Bardeen, ``Theory of Superconductivity'', Handbuch der Physik XV, 274 (1956).
\bibitem{rmp63} B. T. Matthias, T. H. Geballe and V. B. Compton, Rev. Mod. Phys. {\bf 35}, 1 (1963).
\bibitem{rmp64} T. H. Geballe, Rev. Mod. Phys. 36, 134 (1964).
\bibitem{geballe} Theodore H Geballe, Robert H Hammond, and Phillip M Wu, ``What $T_c$ tells'', 
\href{http://www.sciencedirect.com/science/article/pii/S0921453415000362}{this volume}, Physica C (2015).
\bibitem{parks} ``Superconductivity'', ed. by R. D. Parks, Marcel Dekker, Inc, New York, 1969.
\bibitem{little}  W. A. Little, Phys. Rev. {\bf 134}, A1416 (1964). 
\bibitem{ginsburg} V. L. Ginzburg, Phys. Lett. {\bf 13}, 101 (1964).
\bibitem{allender} D. Allender, I. Bray, and J. Bardeen, Phys. Rev. B {\bf 7}, 1020
(1973).

\bibitem{sleight}
A.W. Sleight, J.L. Gillson, and P.E. Bierstedt, ``HighÐTemperature
Superconductivity in the BaPb$_{1-x}$Bi$_x$O$_3$ System'', Solid State Commun., 17 (1975) 27-
28.

\bibitem{matthias_anderson} P.W. Anderson and B.T. Matthias, Science {\bf 144} 373 (1964).
\bibitem{matthias2} B. T. Matthias, ``Anticorrelations in superconductivity'', Physica {\bf 55}, 69 (1971).

\bibitem{stewart} G. R. Stewart, ``Ted Geballe: A lifetime of contributions to superconductivity'', \href{http://www.sciencedirect.com/science/article/pii/S092145341500060X} {this volume}, Physica C (2015).
\bibitem{epilogue} C. W. Chu, P. C. Canfield, R. C. Dynes, Z. Fisk, B. Batlogg, G. Deutscher,  T. H. Geballe, Z. X. Zhao, R. L. Greene, H. Hosono, M. B. Maple, ``Epilogue: Superconducting Materials Past, Present and Future'', \href{http://arxiv.org/abs/1504.02488}{this volume}, Physica C (2015).
\bibitem{c1} G.W. Webb, F. Marsiglio, J.E. Hirsch,   ``Superconductivity in the elements, alloys and simple compounds'',  \href{http://www.sciencedirect.com/science/article/pii/S0921453415000647}{this volume}, Physica C (2015).
\bibitem{c2} G. R. Stewart, ``Superconductivity in the A15 structure'', \href{http://www.sciencedirect.com/science/article/pii/S0921453415000404} {this volume}, Physica C (2015).
\bibitem{c3} E. Bustarret, ``Superconductivity in doped semiconductors'', \href{http://www.sciencedirect.com/science/article/pii/S0921453415000489}{this volume}, Physica C (2015).
\bibitem{c4} Katsuya Shimizu, ``Superconductivity from insulating elements under high pressure'', \href{http://www.sciencedirect.com/science/article/pii/S0921453415000623} {this volume}, Physica C (2015).
\bibitem{c5} Robert P. Smith, Thomas E. Weller, Christopher A. Howard, Mark P.M. Dean, Kaveh C. Rahnejat,
Siddharth S. Saxena, Mark Ellerby, ``Superconductivity in graphite intercalation compounds'', \href{http://www.sciencedirect.com/science/article/pii/S0921453415000568}{this volume}, Physica C (2015).
\bibitem{c6} J.J. Hamlin, ``Superconductivity in the metallic elements at high pressures'', \href{http://www.sciencedirect.com/science/article/pii/S0921453415000593} {this volume}, Physica C (2015).
\bibitem{c7} Viktor V. Struzhkin, ``Superconductivity in compressed hydrogen-rich materials: Pressing
on hydrogen'',  \href{http://www.sciencedirect.com/science/article/pii/S0921453415000441}{this volume}, Physica C (2015).
\bibitem{c8} Richard A. Klemm, ``Pristine and intercalated transition metal dichalcogenide
superconductors'', \href{http://www.sciencedirect.com/science/article/pii/S0921453415000507}{this volume}, Physica C (2015).
\bibitem{c9} Octavio Pe$\tilde{\rm n}$a, ``Chevrel phases: Past, present and future'', \href{http://www.sciencedirect.com/science/article/pii/S0921453415000465}{this volume}, Physica C (2015).
\bibitem{c10}  C.T. Wolowiec, B.D. White, M.B. Maple, ``Conventional magnetic superconductors'', \href{http://www.sciencedirect.com/science/article/pii/S0921453415000908}{this volume}, Physica C (2015).
\bibitem{c11} Yen-Hsiang Lin, J. Nelson, A.M. Goldman, ``Superconductivity of very thin films: The superconductor-insulator
transition'', \href{http://www.sciencedirect.com/science/article/pii/S0921453415000143} {this volume}, Physica C (2015).
\bibitem{c12} Sergey L. Bud'ko, Paul C. Canfield, ``Superconductivity of magnesium diboride'', \href{http://www.sciencedirect.com/science/article/pii/S0921453415000519}{this volume}, Physica C (2015).
\bibitem{p1} Arthur W. Sleight, ``Bismuthates: BaBiO$_3$ and related superconducting phases'', \href{http://www.sciencedirect.com/science/article/pii/S0921453415000398}{this volume}, Physica C (2015).
\bibitem{p2} Arthur P. Ramirez, ``Superconductivity in alkali-doped C$_{60}$'', \href{http://www.sciencedirect.com/science/article/pii/S0921453415000416}
{this volume}, Physica C (2015).
\bibitem{p3} Chandan Mazumdar and R. Nagarajan, ``Quaternary borocarbides: Relatively high T$_c$ intermetallic superconductors and 
magnetic superconductors'', \href{http://www.sciencedirect.com/science/article/pii/S0921453415000611}{this volume}, Physica C (2015).
\bibitem{p4} J.L. Sarrao, , E.D. Bauer, J.N. Mitchell, P.H. Tobash, and J.D. Thompson, ``Superconductivity in plutonium compounds'',
\href{http://www.sciencedirect.com/science/article/pii/S0921453415000581}{this volume}, Physica C (2015).
\bibitem{p5} S. Gariglio, M. Gabay, J. Mannhart, and J.-M. Triscone, ``Interface Superconductivity'',
\href{http://www.sciencedirect.com/science/article/pii/S0921453415000556}{this volume}, Physica C (2015).
\bibitem{p6} Yoshihiro Kubozono, Hidenori Goto, Taihei Jabuchi, Takayoshi Yokoya, Takashi Kambe, Yusuke Sakai, Masanari Izumi, Lu Zheng, Shino Hamao, Huyen L.T. Nguyen, Masafumi Sakata, Tomoko Kagayama, and Katsuya Shimizu, ``Superconductivity in aromatic hydrocarbons'',
\href{http://www.sciencedirect.com/science/article/pii/S0921453415000428}{this volume}, Physica C (2015).
\bibitem{p7} Satoshi Sasaki and Takeshi Mizushima, ``Superconducting doped topological materials'',
\href{http://www.sciencedirect.com/science/article/pii/S0921453415000453}{this volume}, Physica C (2015).
\bibitem{p8} D. Yazici, I. Jeon, B.D. White, and M.B. Maple, ``Superconductivity in Layered BiS$_2$-Based Compounds'',
\href{http://www.sciencedirect.com/science/article/pii/S0921453415000520}{this volume}, Physica C (2015).
\bibitem{p9} Yakov Kopelevich, , Robson R. da Silva, and Bruno C. Camargo, ``Unstable and elusive superconductors'',
\href{http://www.sciencedirect.com/science/article/pii/S0921453415000544}{this volume}, Physica C (2015).
\bibitem{u1} B.D. White, J.D. Thompson, and M.B. Maple, ``Unconventional Superconductivity in Heavy-Fermion Compounds'',
\href{http://www.sciencedirect.com/science/article/pii/S0921453415000714}{this volume}, Physica C (2015).
\bibitem{u2} S.E. Brown, ``Organic superconductors: the Bechgaard salts and relatives'',
\href{http://www.sciencedirect.com/science/article/pii/S092145341500057X}{this volume}, Physica C (2015).
\bibitem{u3} C.W. Chu, L.Z. Deng, and B. Lv, ``Hole-doped cuprate high temperature superconductors'',
\href{http://www.sciencedirect.com/science/article/pii/S0921453415000878}{this volume}, Physica C (2015).
\bibitem{u4} P. Fournier, ``T$^\prime$ and Infinite-Layer Electron-Doped Cuprates'',
\href{http://www.sciencedirect.com/science/article/pii/S0921453415000635}{this volume}, Physica C (2015).
\bibitem{u5} Ying Liu and Zhi-Qiang Mao, ``Unconventional superconductivity in Sr$_2$RuO$_4$'',
\href{http://www.sciencedirect.com/science/article/pii/S0921453415000660}{this volume}, Physica C (2015).
\bibitem{u6} Yuichi Kasahara, Kazuhiko Kuroki, Shoji Yamanaka, and Yasujiro Taguchi, ``Unconventional superconductivity in electron-doped layered 
metal nitride halides MNX (M = Ti, Zr, Hf; X = Cl, Br, I)'', \href{http://www.sciencedirect.com/science/article/pii/S0921453415000490}{this volume}, Physica C (2015).
\bibitem{u7} Andrew D. Huxley, ``Ferromagnetic Superconductors'',
\href{http://www.sciencedirect.com/science/article/pii/S0921453415000532}{this volume}, Physica C (2015).
\bibitem{u8} Hiroya Sakurai, Yoshihiko Ihara, and Kazunori Takada, ``Superconductivity of cobalt oxide hydrate, Na$_x$(H$_3$O)$_z$CoO$_2$$\cdot y$H$_2$O'',
\href{http://www.sciencedirect.com/science/article/pii/S0921453415000374}{this volume}, Physica C (2015).
\bibitem{u9} F. Kneidinger, E. Bauer, I. Zeiringer, P. Rogl, C. Blaas-Schenner, D. Reith, and R. Podloucky, ``Superconductivity in non-centrosymmetric materials'',
\href{http://www.sciencedirect.com/science/article/pii/S092145341500043X}{this volume}, Physica C (2015).
\bibitem{u10} Hideo Hosono and Kazuhiko Kuroki, ``Iron-Based Superconductors: current status of materials and pairing mechanism'',
\href{http://www.sciencedirect.com/science/article/pii/S0921453415000477}{this volume}, Physica C (2015).
\bibitem{u11} C.C. Chang, T.K. Chen, W.C. Lee, P.H. Lin, M.J. Wang, Y.C. Wen, P.M. Wu, and M.K. Wu, ``Superconductivity in Fe-chalcogenides'',
\href{http://www.sciencedirect.com/science/article/pii/S0921453415000386}{this volume}, Physica C (2015).


\bibitem{transient} W. Hu et al, Nature Materials {\bf 13}, 705 (2014).
\bibitem{bozovic} I. Bozovic et al, Jour. of Sup. {\bf 7}, 187 (1994). 
\bibitem{hps}   V. V. Brazhkin,  High Pressure Research: An International Journal  {\bf 27}, 333 (2007).
\bibitem{matthias} B. T. Matthias, Physica {\bf 69},  54 (1973). 
\bibitem{csi} M. I. Eremets, K. Shimizu, T. C. Kobayashi, K. Amaya,  Science {\bf 281}, 1333  (1998).
\bibitem{c6i6} E. Iwasaki et al, Synthetic Metals {\bf 120}, 1003 (2001).
\bibitem{iodanil} T.  Yokota et al, Physica B {\bf 304}, 6 (2001).
\bibitem{struzhkin} See Ref. [\onlinecite{c7}] and references therein.
\ \end{references}
\end{document}